\documentclass[10pt,conference]{IEEEtran}
\IEEEoverridecommandlockouts
\usepackage{cite}
\usepackage{amsmath,amssymb,amsfonts}
\usepackage{algorithmic}
\usepackage{graphicx}
\usepackage{textcomp}
\usepackage{xcolor}[dvipsnames, most]
\usepackage[most]{tcolorbox}
\def\BibTeX{{\rm B\kern-.05em{\sc i\kern-.025em b}\kern-.08em
    T\kern-.1667em\lower.7ex\hbox{E}\kern-.125emX}}
\begin{document}

\title{Investigating the Developer eXperience of LGBTQIAPN+ People in Agile Teams\\
% {\footnotesize \textsuperscript{*}Note: Sub-titles are not captured in Xplore and
% should not be used}
\thanks{This study was financed in part by the Coordenação de Aperfeiçoamento de Pessoal de Nível Superior - Brasil (CAPES) - Finance Code 001}
}

\author{\IEEEauthorblockN{Edvaldo Wassouf-Jr, Pedro Fukuda and Awdren Fontão}
\IEEEauthorblockA{\textit{Faculty of Computing} \\
\textit{Federal University of Mato Grosso do Sul}\\
Campo Grande, Brazil \\
edvaldo.junior@ufms.br, pedro.fukuda@ufms.br, awdren.fontao@ufms.br}
}

\maketitle

\begin{abstract}

Diversity in software teams drives innovation and enhances performance, but it also introduces challenges that require intentional management. LGBTQIAPN+ professionals in the software industry face unique barriers, including discrimination, low visibility, and harassment, which can diminish satisfaction, productivity, and retention. This study investigates the Developer Experience (DX) of LGBTQIAPN+ individuals in Agile software development teams through a survey of 40 participants. Findings highlight that psychological safety and inclusive policies are critical for fostering equitable contributions and team cohesion. Agile practices, such as retrospectives, pair programming, and daily meetings, enhance collaboration and reduce biases when tailored to the needs of underrepresented groups, creating an environment of mutual respect and openness. Additionally, remote work offers significant benefits for LGBTQIAPN+ professionals, including improved psychological comfort, productivity, and work-life balance. However, challenges like isolation and insufficient virtual team interactions remain and must be addressed. This research underscores the importance of integrating inclusivity into Agile methodologies and organizational practices to support the unique needs of diverse professionals. By fostering an environment that values diversity, organizations can enable more effective and satisfied teams, ultimately driving higher-quality outcomes and improved organizational performance. This study provides actionable insights for creating more inclusive and supportive Agile work environments.
\end{abstract}


\section{Introduction}

The software industry recognizes the value of diversity, as diverse teams provide significant advantages for companies \cite{Grundy2024}. Although differences can lead to conflicts and challenges, the benefits of diversity often outweigh these issues \cite{de2023benefits, hoffmann2022human, prikladnicki2005aspectos, miranda2020towards}. The opportunities for improvement in team management can be identified to enhance the retention of professionals \cite{de2023benefits}, since developer satisfaction directly contributes to increased daily productivity \cite{hoffmann2022human, prikladnicki2005aspectos, miranda2020towards}.

In agile software development teams, trust is essential. Communication and context form the foundation of effective collaboration, and a lack of trust can lead to irreversible consequences, such as reluctance to share information and a decline in product quality \cite{de2023benefits, silveira2019systematic}. From this perspective, one of the core principles of the Agile Manifesto is \textit{\textbf{Individuals and Interactions Over Processes and Tools}}, which emphasizes the importance of constructing projects around motivated individuals. Additionally, professionals must receive adequate support to foster confidence in their ability to perform their tasks effectively \cite{prikladnicki2005aspectos, fowler2001agile}.

Dissatisfied software professionals often perceive their productivity as lower than their potential, leading to regrets about the quality of the products generated by their work. Moreover, individuals who are unhappy with their outputs and roles may make work-related decisions that negatively affect software quality and the delivery of value to customers \cite{girardi2021emotions, juarez2021covid, borg2024requirements, graziotin2014happy, graziotin2017consequences, fagerholm2012developer, wazlawick2019engenharia}.

Discrimination and invisibility \cite{de2023post}, low perceived diversity, and harassment \cite{poncell2022diversity} are factors that adversely impact the experiences of LGBTQIAPN+ professionals in the software industry \cite{de2023benefits}. Current research underscores the importance of research focused on the LGBTQIAPN+ population (Lesbians, Gays, Bisexuals, Transgender individuals, and other groups who do not conform to traditional gender and sexual norms) within the software industry \cite{de2023benefits} and in Software Engineering \cite{boman2024breaking}. Research conducted by Ford et al. \cite{ford2019remote} emphasizes the need to develop practices that enhance LGBTQIAPN+ visibility among employees in technology companies. 

 In this context, it is crucial to investigate the aspects of Developer Experience (DX) for this population to identify solutions that can enhance their satisfaction. As noted by Silveira and Prikladnicki \cite{silveira2019systematic} in a systematic mapping study, "\textit{There are research studies about Diversity in Software Engineering, but the literature is missing papers on how Diversity impacts Agile Methodologies.}". In this study, we conducted an interpretive opinion survey with 40 participants to answer "\textit{What are the perceptions of LGBTQIAPN+ software developers about their DX within agile teams?}". We discovered evidence of:
\begin{itemize}
    \item LGBTQIAPN+ professionals thrive in teams where diversity is actively embraced, psychological safety is ensured, and discriminatory behaviors are swiftly addressed. Organizations should develop inclusive policies and provide training to support diverse team dynamics, enabling all professionals to contribute effectively;
    \item Agile practices such as retrospectives, pair programming, and daily meetings are effective in enhancing team collaboration and mitigating biases. These practices should be tailored to address challenges faced by underrepresented groups, creating a culture of mutual respect and openness;
    \item Remote work models demonstrate significant benefits, including improved productivity, psychological comfort, and work-life balance for LGBTQIAPN+ professionals. However, organizations must also address challenges like isolation by fostering strong virtual team interactions and ensuring an inclusive environment in hybrid and on-site settings.
\end{itemize}

\section{Background and Related Work}

\subsection{Developer Experience (DX)}

Fagerholm and Münch (2012)\cite{fagerholm2012developer} define Developer Experience (DX) as a concept that captures developers' perceptions—how they think and feel about their activities within the work environment, their teams, and the software development processes they engage in. Enhancements in DX can have a positive impact on the outcomes of software development projects. Therefore, promoting improvements in the DX of software professionals is essential for delivering value \cite{klotins2023continuous, greiler2022actionable, kropp2020satisfaction}.

Furthermore, research \cite{dutra2021human, machuca2022perceptions, prikladnicki2005aspectos} has highlighted the influence of non-technical aspects, including human and social factors, on productivity and value delivery in agile software development. Among these factors, communication, collaboration, knowledge sharing, and motivation significantly impact team dynamics.

There is an emerging body of research in the literature that identifies various factors influencing DX, often referred to as DX Factors. These factors can be categorized into three dimensions: conation, affect, and cognition \cite{greiler2022actionable, fagerholm2012developer, d2024measuring}. A systematic review of the literature (SLR) conducted by Razzaq \cite{razzaq2024systematic} discusses elements that affect and improve DX. Among the motivations and DX factors highlighted, non-technical aspects such as psychological safety and psychological distress are noted as significant influences on the experiences of professionals in the field.

\subsection{Diversity and Inclusion in Software Engineering} 

Perceived diversity is a concept that helps researchers examine how low-diversity environments impact the experiences of professionals in the industry \cite{rodriguez2021perceived, poncell2022diversity}. This concept can be utilized to gather evidence regarding diversity in the software industry and to assess its effects on professional satisfaction within teams and organizations.

Environments lacking in diversity can perpetuate gender and sexual biases that negatively affect underrepresented populations, such as the LGBTQIAPN+ community. In the software industry, the presence of unconscious biases can hinder the inclusion, participation, and productivity of these underrepresented groups \cite{de2020diversity, prana2021including}. Research has demonstrated that unconscious gender biases can undermine women’s contributions and participation in teams, adversely affecting their overall experience and inclusion in the software industry \cite{duran2024exploring, imtiaz2019investigating, kanij2024enhancing, trinkenreich2022women}.

This scenario underscores the obstacles present in the software industry, particularly the challenges of including underrepresented groups due to hiring preferences in a market dominated by white male workers \cite{weisshaar2024hiring, hussain2020human, trinkenreich2022empirical, campero2021hiring}. Consequently, research on diversity and inclusion is gaining traction by focusing on the underrepresentation of minority groups, particularly regarding gender and race \cite{rodriguez2021perceived, canedo2021breaking, van2023still, trinkenreich2022empirical, gama2024much, dagan2023building, albusays2021diversity, sanchez2021framework, richard2021effects, gunay2020improving, verwijs2023double}.

Moreover, since the onset of the Covid-19 pandemic, the inclusion of minority groups, including the LGBTQIAPN+ population, has led to the formation of more diverse teams \cite{santos2024exploring, ezeilo2023coronovirus, de2023benefits}. As a result, new challenges regarding the inclusion of this population in agile teams have become a focal point for researchers. This community is often marginalized and historically discriminated against, which adversely affects their inclusion in formal employment and retention within the industry \cite{de2023post, de2023benefits}.

\subsection{Work models}
\subsubsection{Remote Work}
In a study carried out on job offers in startups, the authors report that the offer of remote jobs attracts more experienced and diverse candidates, belonging to underrepresented minority groups \cite{hsu2024remote}.
 In this context, gender, racial, and sexual diversity is the subject of analysis for current research \cite{de2023benefits}, \cite{ford2019remote}, \cite{nicholson2022remote}.
 
 This reality allows teams with diverse compositions to be composed of LGBTQIAPN+ professionals. A study conducted by de Souza-Santos et al. \cite{de2023benefits} points out that the benefits (control of identity, identity sharing control, psychological and physical safety) outweigh the limitations (invisibility and isolation) that involve the experience of these software professionals.
\subsubsection{Onsite and Hybrid Work}
In the post-pandemic scenario, studies have focused on assessing the resilience and adaptation of professionals in hybrid work environments. Research demonstrates the difficulties \cite{li2024you}, \cite{de2023post}, \cite{nicholson2022remote} faced by underrepresented populations and highlights the need for support from corporations to maintain and promote diverse teams.

The difficulties faced by the LGBTQIAPN+ population in the software industry include sexual, moral and psychological harassment, discrimination and isolation, as well as the fear of physical violence in the workplace. Thus, these factors become stressors and can affect the experience of these professionals, especially transgender professionals \cite{ford2019remote},\cite{poncell2022diversity},\cite{de2023post},\cite{de2020diversity},\cite{nicholson2022remote}.

\subsection{Developer Experience (DX) of LGTQIAPN+ people}

In the daily operations of software developers, the importance of culture and collaboration is crucial, as these elements are fundamental to DX. They have a direct impact on both productivity and well-being. Key components \cite{greiler2022actionable} include support from colleagues, occasional frustrations, the connections among team members, the use of agile collaboration practices, and the assistance provided in managing the demands and workflows that come with team dynamics.

Opinion surveys support assessing inclusion in software engineering \cite{johnson2024strategies}. They offer valuable insights into underrepresented communities, including women, ethnic minorities, gender-diverse groups, and neurodivergent individuals. To achieve this, it is important to utilize tools that effectively capture subjective factors such as "flow," where developers are fully engaged, and "focus," which relates to maintaining concentration on tasks \cite{d2024measuring}.

Greiler \cite{greiler2022actionable} presents a framework for understanding and improving the developer experience (DX) of software professionals. Among the DX factors identified by the authors are support, feeling connected, collaboration and culture, and having aligned values. In this context, using surveys can be an effective strategy to capture these elements and evaluate DX from the perspective of specific contexts and cultures, within underrepresented developers, such as the LGBTQIAPN+ community.

\section{Research Method}

\subsection{Goal and Research Questions}
We performed an interpretive opinion survey with a quantitative and qualitative paradigm following the guidelines described by Mollieri et al. \cite{molleri2016survey}. The GQM \cite{basili1994gqm} was used to construct the research questions to capture elements about the developer experience of LGBTQIAPN+ people in agile teams.

\textbf{\textit{RQ: What are the perceptions of LGBTQIAPN+ software developers about their DX within agile teams?}} \underline{Rationale:}  We aim to investigate the specific perceptions of LGBTQIAPN+ software developers regarding their DX within agile teams, given the unique challenges they face, such as discrimination, psychological discomfort, and inadequate inclusion policies \cite{de2023benefits}. Agile methodologies emphasize collaboration and team dynamics, making it crucial to understand how these factors intersect with the experiences of underrepresented groups.

To answer this research question, we developed auxiliary questions to capture the factors influencing DX based on the work of Greiler \cite{greiler2022actionable} and Fagerholm and Munch \cite{fagerholm2012developer}. We used the influence factors categorized into three dimensions (affect - how developers feel about their work; conation - How the developers see their values embodied in experiencing some objects or
processes/methods/activities they perform; cognition - refers to how developers perceive objects, such as tools, techniques, technical environment) by Razzaq et al. \cite{razzaq2024systematic} (\textbf{Table 1}).

\begin{itemize}
    \item  \textit{A1}: What is the perception of LGBTQIAPN+  developers regarding growth opportunities within the company and the factors that influence their career trajectories?
    \item  \textit{A2}: What are the main factors affecting the engagement and perception of LGBTQIAPN+ developers about their teams and work processes?
    \item  \textit{A3}: How do agile practices and different work models influence the DX of LGBTQIAPN+ developers in the corporate environment?
\end{itemize}

\begin{table}[!htbp]
\caption{Research Questions - DX Factors}
\centering
\begin{tabular}{|l|l|l|l|} 
\hline
\textbf{RQ} & \textbf{DX Factors}~ & \textbf{Description} & \textbf{Dimension} \\ 
\hline
A1 & Motivation; & \begin{tabular}[c]{@{}l@{}}Career opportunities, \\Working conditions,\\Participation;\end{tabular} & \begin{tabular}[c]{@{}l@{}}Conation,\\affect;\end{tabular} \\ 
\hline
A2 & \begin{tabular}[c]{@{}l@{}}Motivation,\\Team Work,\\Developer\\Attributes;\end{tabular} & \begin{tabular}[c]{@{}l@{}}Defined team culture,Team \\Collaboration, Team maturity, \\Supportive relationships, \\Team~structure and\\more expertise, Conflicts/\\congruence, Psychological\\distress,~Avoiding \\Collaborating Conflicts, Pair\\Programming,~Sprint\\Planning Sessions,\\Psychological Safety;\end{tabular} & \begin{tabular}[c]{@{}l@{}}Conation, \\Affect,\\Cognition;\end{tabular} \\ 
\hline
A3~ & \begin{tabular}[c]{@{}l@{}}Motivation,\\Team Culture;\end{tabular} & \begin{tabular}[c]{@{}l@{}}Work-Life-Balance, \\Working conditions, \\Global distance, \\Psychological safety;\end{tabular} & \begin{tabular}[c]{@{}l@{}}Conation,\\Affect;\end{tabular} \\
\hline
\end{tabular}
\end{table}

\section{Survey Design}

\subsection{Instrument Design}

We developed the survey following the recommendations in Mollieri et al.\cite{molleri2016survey}. Initially, an early version of the survey was created and iteratively refined. This version was then subjected to a pilot test with six participants, including researchers and postgraduate students. Based on their feedback, the adjustments were made to build the current survey version.

The survey\footnote{bit.ly/3YiMrQx} consisted of two types of questions: quantitative and qualitative. The quantitative questions included single-choice options and questions using a Likert scale. The qualitative component was made up of open-ended questions. Each section of the form contained both types of questions. To ensure participants' comfort, they were allowed to skip questions that did not apply to their professional context, avoiding forced responses.

The first page of the survey provided a summary of the research along with a consent form for participants to review. The \textbf{\textit{first two sections}} cover questions aimed at characterizing participants based on their demographic information (seniority, gender identity, sexuality, and
work model). \textbf{\textit{The third section}} focused on the characteristics of the team in which the participants either currently work/worked (i.e., Team categories / Team formation phase) as follows: 

\subsubsection{Team categories}\label{AA}

To categorize the teams for subsequent data analysis, two classifications were used, present in \cite{katzenbach2001equipes, liboreiro2018gestao, montanari2011maturidade, santos2014social}. The first category was intended to classify the team's performance. Otherwise, the second aimed to categorize the growth state of the team. The first one provides five classifications: \textit{Workgroup}: In this group, each person has individual responsibilities and objectives and they do not yet identify a reason for being a team. Absence of collective performance requirements. \textit{Pseudo-team}: This group have the worst performance, as individual performance is highlighted, and the results obtained together are inferior to individual performance. \textit{Potential team}: This group works together on its deliveries, but its members need to understand its purpose, its objectives, its products, and its tasks. \textit{Real team}: A real team is made up of people with complementary skills and committed to each other through a common goal and well-defined work approaches. \textit{High-performance team}: A high-performance team, in addition to having all the requirements of a real team, its members are committed to the personal growth and success of each team member.

\subsubsection{Team formation phase}
To classify the team according to the team's current formation phase, the following categories were used: \textit{Formation}: This is the initial formation of the team, in which members are beginning to interact. It is characterized by feelings of insecurity and uncertainty regarding the group’s goals, structure, and leadership. \textit{Confusion/Conflict}: it is characterized by the occurrence of various conflicts in the group. It is a period of confrontation, disunity, tension, and hostility. \textit{Normalization}: Cohesion begins to emerge in the group’s behavior, bringing members closer to one another. At this stage, the team develops basic rules or norms for working together. \textit{Performance}: the stage in which the group structure is functional and accepted. At this stage, the group is cohesive, and its energy is focused on the tasks. \textit{Disintegration}: the final stage of group development. Since the activities must be completed and the group dissolved, the focus is no longer on task performance but on completing the work.

The \textbf{\textit{fourth section}} addressed the company's approach to handling bias and how this approach impacts employees' experiences. The \textbf{\textit{fifth section}} explored the participant's relationship within their team. The \textbf{\textit{sixth session section}} asks participants about recommendations for agile practices that help reduce barriers, capturing professionals' opinions on recommendations to reduce bias in the job market.

\subsection{Participants}

Regarding this point, we present our adherence to reporting and conducting sampling recommended by Baltes and Ralph \cite{baltesandralph}:

\textbf{Philosophical position.} We employed the interpretivism paradigm that focuses on understanding social phenomena from the perspective of individuals, aiming to interpret their experiences, meanings, and contexts.

\textbf{The purpose of sampling.} To address challenges associated with sampling hidden or underrepresented populations, we adopted a respondent-driven sampling (RDS) strategy as outlined by Baltes and Ralph \cite{baltesandralph} and Santos and Gama \cite{de2024hidden}. RDS mitigates traditional sampling biases by initiating recruitment with diverse "seeds" from the target population and leveraging participants' social networks for referrals, with controlled recruitment waves to limit the overrepresentation of highly connected individuals. It allowed us to sample to obtain initial data, generate hypotheses, or discover new patterns, and not with the goal of generalization.

\textbf{Selecting our sample. }The survey was distributed during November 2022 and April 2024. It was distributed through direct email invitations to eligible participants as well as shared across software development communities on LinkedIn, Twitter, and the Dev Community\footnote{https://dev.to/}.  Additionally, we reached out to researchers in the field who specialize in the human aspects of software engineering (SE), and they assisted in distributing the survey within their professional networks. Additionally, researchers in the field who work on issues related to the human aspects present in SE were contacted and helped distribute the form in their professional networks.

With these recommendations, we reached out to 43 participants completing the survey, resulting in a response rate of 22\%. Three responses were discarded after demographic checks, we identified that the participants were heterosexual and had no identification with the LGBTQIAPN+ community. 

\section{Data analysis}

For data analysis, the data were prepared in spreadsheets\footnote{https://figshare.com/s/4ba1a4048cc17148a1ba} and split into quantitative and qualitative data. From this, for quantitative sampling analysis, subgroups were constructed according to sexual orientation, gender, category and status of the team, and work model.

To analyze the \textbf{\textit{quantitative}} data collected, we calculated the overall agreement for the general group by averaging the agreement of all participants on each issue. Subsequently, we compared the agreement levels across various subgroups, which were formed based on the participants' demographics and team categorizations. For the questions using a 5-point Likert scale, we categorized responses into levels of disagreement (partial disagreement + strong disagreement) and levels of agreement (strong agreement + agreement) to represent the general tendencies towards disagreement or agreement.

For the \textbf{\textit{qualitative}} analysis, open-ended questions allowed us to explore participants' justifications for their choices on simpler questions, as well as to gather evidence about their satisfaction and perceptions regarding their experiences within the team and the company. The thematic synthesis methodology was used following the recommendations presented by Cruzes and Dyba \cite{cruzes2011recommended}. An example of the processing of data extracted from participants is shown in \textbf{Table III}.

The process of extracting keywords and excerpts of interest, open coding, and extracting subthemes was carried out manually. For theme extraction, we used the support of an LLM (Large Language Model) tool (ChatGPT-4) following the recommendations in \cite{roberts2024artificial, yan2024human} which is used as an efficient tool for theme extraction. The prompt used in this study is based on "structured task description" and "Input-Process-Output (IPO)" patterns. It is described below.

\begin{tcolorbox}[colback=blue!10, colframe=blue!50]
\footnotesize
I am performing a thematic synthesis process based on the responses to a survey. I will provide the questions, subthemes, and manually extracted codes, then I will ask you to generate possible themes from the codes and subthemes.

Question (Q1) - Description of the question;

Associated codes - Sequence of manually extracted codes;

Subthemes - Sequence of subthemes;
\end{tcolorbox}

% \section{Results and Discussion}
% In this section we present the main results of the survey and discuss the triangulation of these data according to the answers to the auxiliary questions. For the quantitative analysis the main results of the general group and differences that differ from the subgroups are presented, in addition, the data from the extraction of the thematic synthesis are discussed.

\section{Results}
This section presents the main findings of the quantitative and qualitative questions that help answer the auxiliary research questions. Regarding the \textbf{\textit{Demographics}}, the sample of participants showed male dominance, a fact that is present even in this underrepresented group. Most professionals work remotely (70\%) and there was little participation in Gender Nonconforming (GNC - transgender/non-binary) groups,  only four professionals indicated that they were part of this population. This data is shown in the table \textbf{Table II}. 

The qualitative results presented were obtained through thematic synthesis. To address the auxiliary research questions, the findings are organized according to the open-ended questions in the questionnaire\footnote{bit.ly/3YiMrQx}. Corresponding codes identify relevant excerpts from participants' responses. Each respondent is designated by an identifier (P) followed by a number indicating the sequence of their response.

\begin{table}[ht!]
\centering
\caption{Distribution of participants by gender identity, sexuality, work experience, position, work model, team formation phase, team performance.}
\label{tab:minha_tabela}
\begin{tabular}{llr}
participants &  &  Percentage \% \\ \hline
Gender Identity & Cisgender Male & 60\\
& Cisgender Female & 30 \\ 
& Transgender Male & 2.5 \\
& Transgender Female & 2.5 \\
& Non-binary & 5 \\ \hline
& Gay & 45\\
Sexuality & Lesbian & 17.5\\
& Bisexual & 25 \\ 
& Pansexual & 7.5 \\
& Asexual & 5 \\ \hline
Seniority & Senior & 37.5\\
& Junior & 35 \\
& Mid Level & 22.5 \\ 
& Not informed & 5 \\ \hline
Position & Developer & 50\\
& Product Owner & 12.5 \\ 
& DevOps & 7.5 \\
& Quality Assurance & 7.5 \\
& Product Designer & 5 \\
& UI Designer & 5 \\
& Project Manager & 2.5 \\
& Tech Lead & 2.5 \\
& Team Lead & 2.5 \\
& Software Architect & 2.5\\
& Stakeholder & 2.5 \\ \hline
Work Model& Remote & 70 \\
 & On-site & 15 \\
 & Hybrid  & 15 \\ \hline
 Team's \textbf{formation} phase& Performance & 45 \\
 & Normalization & 35 \\
 & Formation  & 10 \\ 
 & Confusion/Conflict  & 10 \\ \hline
 Team’s \textbf{performance} & Real Team & 42.5 \\
 & Potential Team & 25 \\
 & High Performance Team  & 15 \\ 
 & WorkGroup  & 7.5 \\
 & Pseudo-Team  & 7.5 \\ 
 & Not apply & 2.5\\ \hline
 
\end{tabular}
\end{table}

\subsection{(A1) What is the perception of LGBTQIAPN+ professionals regarding growth opportunities within the company and the factors that influence their career trajectories?}

The perceptions regarding professional growth opportunities within the company reveal a generally favorable outlook among participants. 64.1\% (25) of the group agreed, indicating a positive view, while 25.6\% (10) disagreed, and 10.3\% (4) remained neutral. Notably, members of real teams demonstrated an even stronger agreement rate of 70.5\% (12), with only 17.7\% (3) disagreeing. In comparison, potential team members showed a lower agreement rate of 66.6\% (6) and a higher disagreement rate of 33.3\% (3). It can suggest that established teams may provide clearer professional development pathways compared to those still forming.

The participants perceive the alignment of interests with the company they are part of as positive. 62.5\% (25) agreeing and only 20\% (8) disagreeing; 17.5\% (7) indicating neutrality. Among real team members, the agreement rate was slightly higher at 70.58\% (12), and only 5.88\% (1) expressed disagreement, although a notable 23.52\% (4) remained neutral. This indicates that while there is a consensus about alignment with company interests, the higher neutrality among real team members suggests that further efforts may be necessary to strengthen this alignment and ensure all team members feel equally connected to the company’s goals.

The professional journey for the participants has been marked by a welcoming atmosphere and increased opportunities. Diversity has increasingly been recognized as a crucial market strategy, driving improvements in acceptance, reducing conflicts, and enhancing the evolution of the work environment. This shift has fostered the development of inclusive policies and a culture that promotes respect for diversity.
P13 reports a positive view of the industry being open to diversity\textit{``I believe that the software market is quite open to LGBTQIAP+ people. Opportunities are equal regardless of sexual orientation."}. Professionals highlight the company's culture as a central point in this issue, as in the report of P29: \textit{``In general, it is challenging, but it depends exclusively on the company. In my case, the company where I currently work has a very inclusive policy for LGBTQIAP+, which makes the environment more relaxed and welcoming. I have not had any problems regarding my orientation and have always been well received by the teams I have worked for, but this is due to the culture of the company where I work."}.

In contrast, there are frustrations due to the dominance of heterosexual male culture within the workplace. It leads to uncomfortable situations involving integration during agile ceremonies, low tolerance for mistakes, and experiences of jokes and discrimination. These challenges are denoted by P21: \textit{``Full of “obstacles” (prejudices beyond the obstacles of the area itself and many jokes disguised as pranks), and always facing distrust of its capacity and competence."} also in P2: \textit{``...more challenges than heterosexual cis men in their daily lives. The moments when these challenges are most pronounced are in planning and discussing functionality. It is much more common for there to be unfounded disagreements, and mistakes are much less acceptable."}. Participants also noted inequalities in hiring practices, work overload, and fears of social or professional retaliation. P16 captured this toxic environment, stating: \textit{``...in the last company before, heavy atmosphere mainly in calls transphobic, xenophobic jokes and laughter against diversity and inclusion."}.

\subsection{(A2) What are the main factors affecting the engagement and perception of LGBTQIAPN+ professionals about their teams and work processes?}

Regarding team classification, 42.5\% (17) participants identified as belonging to a Real Team, while 25\% (10) categorized themselves as part of a Potential Team. 15\% (6) reported being in a High-Performance Team, and 15\% (6) indicated they were part of either a Working Group or a Pseudo Team. 

Concerning the team's development phase, 45\% (18) of participants reported being in the Performance Phase, suggesting that many teams have reached a level of effectiveness characterized by cohesive structures and a focus on task completion. Meanwhile, 35\% (14) were in the Normalization Phase, indicating ongoing adjustments and the establishment of collaborative norms. However, 10\% (4) indicated being in the Formation Phase or experiencing Confusion/Conflict, highlighting that a minority of teams are still grappling with initial stages of development that involve uncertainty and potential discord. 

Positive experiences related to team characteristics are closely tied to several key factors that contribute to an effective team environment. Participants highlighted the importance of identifying issues during retrospectives, achieving efficiency through well-structured planning, and leveraging new technologies that complement the team's expertise. Other critical aspects include team commitment, mutual respect, internal support, collaborative task execution, diverse skill sets, trust in overcoming challenges, and fostering professional growth. Together, these elements enhance teamwork and strengthen the team's overall performance. P17 reports the benefits of diversity and collaboration: \textit{``My team has diverse skills and we can rely on each other to make up for our shortcomings."}.

Positive experiences related to the team phase are characterized by supportive and collaborative factors. Participants explored the smooth integration of new members, efforts to improve processes, and the absence of fear when admitting technical debt, regardless of seniority. Other key aspects included mutual collaboration towards a common goal, clear hierarchy and alignment, a solid team foundation, and good relationships among team members. These elements foster a productive and cohesive working environment, allowing teams to function more effectively. An example of this is in P5's report: \textit{``...consider my team as real, we support each other in what we need and fight for the same goal. Each person has a skill that the other does not have or has to a lesser degree, we complement each other..."}

Frustrations about team dynamics, however, were connected to more negative aspects such as coercion, poor management, disorganization, individualism, gender bias, collaboration difficulties, task isolation, and limited communication within the team. Participants from less structured teams highlighted these challenges, as reflected in P4’s statement: \textit{``Employee performance is great, but poor management is not limited to coercion or negative feedback, it is also disorganized, which means employees do not know what they have to do or how they have to do it, and even so, what has to be done needs to be and is done."}. This sentiment underscores how poor management practices can create confusion, yet tasks are still completed despite the disarray.

Frustrations about the team phase were tied to negative dynamics, such as disputes over voice and power in decision-making, fear within the decision process, challenges of centralized decision-making, and unresolved technical debt. Some participants, like P4, pointed to the lack of structure as a source of dissatisfaction: \textit{``There is usually a dispute over voice or power over decisions regarding the specified products. Developers and designers are united, as fear allows, but there is no single person who makes decisions."}. Additionally, frustration due to the immaturity of the team was noted by P38, who expressed concern over the team's lack of process knowledge: \textit{``I think the main problem with my team is the lack of knowledge of the processes and formal knowledge of how a software team works."}. 

Recent team changes were marked by continuous adaptations, as highlighted by participants. These changes included team growth, the integration of new members, and adjustments to evolving group dynamics. Despite these transformations, there was a sense of unity as teams worked through the challenges of adaptation, fostering an environment of collaboration and resilience. In terms of additional effort (Q33), participants expressed that gender dynamics and subtle biases required them to exert extra energy to be heard, particularly in Agile ceremonies. Codes such as tone policing, self-policing due to gender differences, and caution in communication were recurrent themes. As P21 described: \textit{``I try to be as objective as possible and set an example to avoid misinterpretations and misunderstandings, and I feel that this is not mutual... I already make an effort to avoid comments and 'providing ammunition' as a pretext for prejudiced and disrespectful jokes, in addition to losing credibility."}.

However, some participants, such as P23, indicated no relation between gender issues and additional effort. These individuals managed their identity by carefully separating their personal and professional lives. As P23 explained: \textit{``Not much, but maybe it's because I've always avoided separating my personal and professional life and I talk little or nothing about my sexuality in a professional environment."}. This approach allowed them to navigate the workplace with more ease, minimizing the need for additional effort to manage bias.

Regarding the influence of gender and sexuality on software engineering processes it reveals significant insights into team dynamics and inclusivity. Concerning gender influencing team demands, the general group showed a notable disagreement with 62.5\% (25) indicating they did not believe gender impacts team requirements. However, the sentiment differed among various team classifications. For instance, 66.66\% (2) of pseudo-team members agreed that gender influences demands, suggesting that less structured teams may be more susceptible to gender biases. Additionally, transgender and non-binary professionals reported mixed responses, indicating a potential sensitivity to gender biases that could impact their experiences in the workplace.

When exploring the influence of sexuality on team demands, a similar pattern emerged. Only 12.5\% (5) of the general group agreed that sexuality influences team requirements, with a substantial 72.5\% (29) disagreeing. Interestingly, among professionals identifying as women, 69.2\% (9) agreed, contrasting sharply with the responses of men, where only 12\% (3) agreed. This disparity suggests that perceptions of sexuality's influence may vary based on gender identity, indicating that women may feel more impacted by sexual orientation in team settings. 

When asked if they consider changing projects due to experiences of sexual or gender discrimination, the responses further underscored the experiences of women and transgender professionals. For instance, among pseudo team members, 66.66\% (2) indicated they would consider changing projects due to discrimination, highlighting the adverse effects of less cohesive team structures on marginalized professionals. In contrast, 83.33\% (5) of high-performance team members responded negatively, suggesting that supportive teams may mitigate the negative impacts of discrimination.

In terms of identity-sharing control, participants reported managing their identity by keeping their sexuality unexposed in the workplace. This approach allowed them to navigate professional settings without drawing attention to their sexual identity, As P28 describes: \textit{``I have always been a little afraid to talk about my sexuality and dating in the workplace. Even though the team members never made prejudiced comments, I was afraid to expose myself, especially because my team is all men and all are straight..."}. Frustration when sharing control was primarily linked to gender bias and the need for more efficient communication, particularly in environments where masculine language dominated. Some participants experienced discomfort due to discriminatory jokes or biased comments, which affected team dynamics. As P4 described: \textit{``Organization, tact, and care for the team or demands usually fall to the female figure whenever possible. Patriarchal inheritance of women's roles as secretaries, nurses, or housewives (who put everything in order, in addition to the house, the man's emotions and psychology)."}. 

On the positive side of sharing control, some participants recognized that their sexuality could contribute to better communication and understanding in software processes, particularly by fostering inclusivity and empathy within teams, exemplified in the account of P22 \textit{``I opened up to the team in a relaxed conversation, saying that my boyfriend was also a developer. The team received the information naturally and never questioned me about anything. On the contrary, I got closer to them and became true friends with two people, in addition to my manager."}. 
 
The engagement of LGBTQIAPN+ professionals within teams is influenced by factors related to performance, tone adjustment, and team belonging. Concerning the impact of jokes and sexual discrimination on performance,  25\% (10) participants agreed that such issues negatively affected their performance, while a notable 60\% (24) disagreed, suggesting a prevailing belief that these factors do not significantly hinder productivity. However, there were important differences across demographic lines; for instance, 53.8\% (7) of women (cisgender and transgender) acknowledged a negative impact, contrasting sharply with 24\% (6) of men (cisgender and transgender). Junior professionals appeared more sensitive to these issues, with 28.57\% (4) agreeing compared to only 6.66\% (1) among senior professionals, highlighting a potential gap in experience and exposure to workplace discrimination.

Among the participants, 40\% (16) believed that tone adjustments were necessary, indicating perceived pressure to modify their communication style based on team dynamics. This perception varied notably in high-performance teams, where 83.33\% (5) disagreed with the need for tone adjustment, suggesting that such teams may cultivate a more confident and cohesive atmosphere. In contrast, 66.66\% (2) of professionals in workgroups agreed with the necessity of tone adjustments, pointing to potential communication challenges in less structured environments. This discrepancy emphasizes the varying demands placed on professionals depending on their team context. Among the participants identifying as women, 38.5\% (5) acknowledged the need for tone adjustments, while 41.7\% (10) of those identifying as homosexual expressed similar sentiments. The data suggest that communication styles and the pressure to conform to team norms significantly impact the engagement of LGBTQIAPN+ professionals. 

In cases where performance was perceived as unaffected, participants credited the absence of discriminatory comments and the existence of a respectful work environment. Factors such as identity-sharing control, respect among all employees, and the possibility of reporting and punishing aggressors were highlighted. HR efforts, policies against sexual and gender discrimination, a safe work environment, and a clear establishment of boundaries further contributed to this positive perception.

Frustrations that affect performance emerged when prejudiced comments, jokes, and attacks on integrity occurred, particularly affecting LGBTQIAPN+ professionals. These incidents often involved transphobia, sexism, and inappropriate behavior from management or customers, leading to a decrease in performance and uncomfortable situations. Low perceived diversity and limited efforts to hire minorities were additional sources of dissatisfaction. As P35 recounted: \textit{``Recently, I suffered a type of transphobic attack from one of the people on the team. I received support from my leader, who was able to act quickly and try to resolve the situation. But it destabilized me and affected my progress."}. This underscores how such incidents can deeply impact individuals and hinder their performance, even with supportive leadership.

% \section{Qualitative Results}
  \begin{table*}
\caption{Thematic synthesis process}
\centering
\begin{tabular}{|l|l|l|l|l|} 
\hline
\multicolumn{5}{|l|}{\textbf{Can you tell us how you consider the career journey of an LGBTQIAPN+ professional in software projects?}}                                                                                                                                                                                                                                                                                                                                                                                                                                                                                                                                               \\ 
\hline
\textbf{Response}                                                                                                                                                                                                                                         & \textbf{Keyword/important excerpts}                       & \textbf{Codes} & \textbf{Subthemes}                          & \textbf{Themes}                                                            \\ 
\hline
\begin{tabular}[c]{@{}l@{}}A constant struggle against stupid jokes and\\~uncomfortable situations.~Sometimes even\\~the audacity to say who I should date or\\~which~~colleague~I make a cute couple with\\~and should stop being gay.\end{tabular}  & \begin{tabular}[c]{@{}l@{}}fight against stupid jokes\\uncomfortable situations.\\.. and should stop being gay.\end{tabular} & \begin{tabular}[c]{@{}l@{}}Prejudiced jokes\\Discomfort\\Psychological \\harassment\end{tabular} & \begin{tabular}[c]{@{}l@{}}Discrimination at \\work\\dissatisfaction\end{tabular}                         & Discrimination                                                              \\ 
\hline
\end{tabular}
\end{table*}

\subsection{(A3) How do agile practices and different work models influence the experience of LGBTQIAPN+ professionals in the corporate environment?}

When discussing positive aspects of in-person work, participants highlighted the benefits of interacting with team members, maintaining a clear separation between personal and professional life, and having more direct communication. On the other hand, dissatisfaction with in-person work was more pronounced, with participants pointing to unproductive environments, the formation of social silos, and exposure to prejudiced behavior such as sexism and misogyny. Some felt psychological discomfort due to these factors, with P12 sharing: \textit{``Unfortunately, I have had the experience of working in a highly sexist and misogynistic environment before, and it was not at all comfortable."}. Additionally, the need for commuting, lack of flexibility, and reduced focus in the office environment contributed to a perception that in-person work was less necessary and more stressful. Conflicts and pressures associated with in-person work were also noted, including micromanagement, excessive demands, and the need for constant self-policing, as P19 stated: \textit{``I've never worked in person... but I always felt very nervous and had to watch myself even more."}. Psychological discomfort and the inability to adapt to noisy, high-pressure environments were common sources of frustration for many professionals.

In terms of positive experiences with remote work, professionals emphasized improved productivity and time management. The flexibility and freedom provided by remote work allowed for better work-life balance, reduced stress, and enhanced overall well-being. participants appreciated not having to commute, experiencing fewer discomforts, and enjoying a safer, more interactive environment. Despite initial challenges and feelings of isolation, satisfaction with remote work remained high, as P21 noted: \textit{``Much more comfortable working, more productive, and with a better quality of life."}. However, dissatisfaction with remote work was also expressed, primarily due to the lack of socialization and feelings of isolation. For those who valued team interaction and collaboration, the absence of in-person contact was a challenge.

Agile practices have been identified as key mechanisms for reducing barriers within teams, with pair programming, retrospectives, and daily meetings playing significant roles. These practices enhance team dynamics and facilitate a whole product vision while fostering a safe agile environment. The need for a robust organizational agile culture is crucial. It involves strengthening team relationships, ensuring positive communication, and establishing respect as a foundational element. Inclusion policies and a culture grounded in Scrum can help reduce prejudice against LGBTQ+ individuals. P9 encapsulated this by stating, \textit{``I consider pair programming, especially if it is with a leader, a great time to improve the relationship and really show your difficulties. But in reality, it is very subjective and depends a lot on each person. No practice will be useful if the leader or any other team member is not approachable, humble and professional."}. 

To promote inclusion and diversity in agile teams, participants discuss the importance of implementing quotas for transgender individuals, providing education on diversity, and offering training for leaders and teams. They emphasized the necessity of normalizing the presence of LGBTQIAPN+ leaders in the workplace. Furthermore, creating anonymous channels for reporting discrimination, implementing swift disciplinary measures, and fostering cultural change within companies are critical steps. As noted in P3, \textit{``I think the simplest thing that can be worked on is the quota system, especially for transgender people, followed by education work on diversity and forms of treatment. Also being stricter when receiving reports of discrimination."}.

\section{Discussion}

We discuss the results from the lens of the agile pillar ``Individuals and Interactions over processes and tools" and DX factors—such as psychological safety, psychological distress, engagement, motivation, developer attributes, software processes, and organizational culture—we compared these findings with our research findings, as well as primary studies focused on the LGBTQIAPN+ community and the factors that influence software professionals’ productivity and engagement. 

\subsection{(A1) Growth Opportunity and Factors that influence career}

The participants' positive outlook on professional growth opportunities and alignment with company interests reflects the importance of fostering effective interpersonal relationships, a core tenet of Agile methodology. The data reveals that established teams exhibit stronger alignment and agreement on professional development compared to potential teams, suggesting that clear communication and strong interpersonal dynamics play a crucial role in the professional growth of individuals. This observation is consistent with Agile's emphasis on the value of individuals and the quality of their interactions on a diversity perspective \cite{silveira2019systematic}.

On the other hand, the frustrations highlighted in the results—such as the dominance of a heterosexual male culture, discrimination, and a lack of tolerance for mistakes—stand in stark contrast to the Agile pillar of prioritizing individuals and interactions over processes and tools. These cultural and systemic issues not only undermine open and respectful communication but also stifle collaboration, both of which are essential for fostering high-functioning Agile teams. Smite et al. \cite{SMITE2021106612} further emphasize how cultural barriers, including the reluctance to expose problems and discuss failure, obstruct an organization's ability to fully embrace agility. Discriminatory behaviors and a toxic work environment prevents trust, stifling the healthy interactions that agile methodologies prioritize. Moreover, the concerns about inequalities in hiring practices, work overload, and fear of retaliation emphasize an organizational focus on hierarchical or procedural concerns at the expense of individual well-being and inclusivity. Sarker \cite{sarker2022identification} argues that developers who experience demotivation and frustration due to harmful interactions—such as verbal abuse, intimidation, or inappropriate behavior from colleagues—may choose to exit an organization .

Regarding career growth opportunity and career path, we found that while there is general agreement between the interests of the professionals and those of the organization, members of less structured teams expressed some disagreement on this topic. On the career's positive experiences: in technology environments where professionals report having a good experience, there tend to be more opportunities and a welcoming environment.

\subsection{(A2) Engagement and perception about their teams and work processes}

Collaboration and increased productivity are often found in more structured teams.
There is already literature discussing team structuring and its relationship with team productivity, such as the work of Riaz et al.\cite{riaz2019effect}. Our approach focused on investigating developers' perceptions of their teams and being able to identify their maturity and performance. When team members feel comfortable discussing technical debt and are committed to improving these challenges, they create a supportive environment that enhances collaboration and teamwork. Situations, where individuals must adjust their tone or hide aspects of their identity (e.g., gender, sexuality), illustrate that a lack of a psychologically safe environment undermines effective collaboration. Agile’s emphasis on “individuals and interactions” implies that every team member should feel comfortable and respected, suggesting these biases directly contradict the principle. 
The literature addresses the impacts of criticism on inclusion, Gunawadena et al. \cite{gunawardena2022destructive} suggest criticism as a factor that impacts the experience of professionals.

Negative experiences like coercion, disorganization, gender bias, and poor communication often stem from breakdowns in human-centered interactions, causing friction and confusion. Even with formal processes, poor interpersonal practices can derail team cohesion, emphasizing that interpersonal respect and clarity outweigh detailed management structures. These challenges, including gender bias, are discussed in works like Poncell and Gama \cite{poncell2022diversity}, Ramos and Gama \cite{de2020diversity}, and Trinkenreich et al. \cite{trinkenreich2021please}, which highlight its impacts, particularly on women. Souza Santos et al. \cite{de2023benefits} also address isolation and fear in SE environments. Furthermore, changes in team dynamics and poor understanding of software processes exacerbate frustrations, as noted by Meyer et al. \cite{meyer2014software} and Ahmad et al. \cite{ahmad2024non}, with such frustrations negatively impacting developers' happiness \cite{graziotin2017unhappy}.

Communication-related frustrations, such as tone policing and gender-based self-policing, can be mitigated through direct communication and boundary-setting between personal and professional life. Gender roles, like team care and work overload, often exacerbate challenges for female professionals. Steinmacher et al. \cite{steinmacher2024breaking} highlight the additional effort women make to be heard in male-dominated environments, while Outão et al. \cite{do2023investigating} reveal how persistent sexism and microaggressions, such as ignoring women's input, act as barriers to their inclusion.

\subsection{(A3) Influence of agile practices and different work models}

Participants highlight the value of identifying issues during retrospectives, giving and receiving mutual support, and fostering diverse skill sets and trust. These practices exemplify placing individuals and interactions at the forefront, as the personal commitment and openness in retrospectives typically drive improvement more effectively than any rigid protocol.

Teams that prioritize continuous agile practices, along with process improvement and planning, tend to foster greater engagement among LGBTQIAPN+ developers. Furthermore, teams with a clear division of tasks tend to achieve more cohesion and alignment with their objectives. This data is consistent with the literature in the works carried out by
 Meyer et al.\cite{meyer2019today}, \cite{meyer2014software} and Fontão et al. \cite{fontao2023developer}.  P9’s comment highlights that the approachability and professionalism of individuals in these interactions are pivotal to the effectiveness of Agile practices, aligning with the emphasis on personal relationships over rigid tools or processes. 

Professionals highlight that the flexibility of remote work when compared to other work models, promotes a better work-life balance. It can lead to increased productivity, environmental control, psychological safety, and an interactive atmosphere while reducing stress. However, remote work is not without its challenges. Many individuals experience feelings of isolation and lack of team interaction, which can create difficulties. This data corroborates the work of Souza Santos et al. \cite{de2023benefits} that explores the limitations of remote work, as well as its advantages for LGBTQIAPN+ professionals. While remote work fosters productivity and flexibility, the reported feelings of isolation highlight a potential gap in interactions. Agile teams rely on frequent and meaningful collaboration, which may require intentional practices to bridge the gap between remote and in-person interaction dynamics.

In-person work also presents its own set of challenges. P12’s and P19’s comments underscore environments where interactions are harmful or stress-inducing, which directly contrasts with Agile's emphasis on nurturing positive and meaningful human relationships. Professionals may face issues such as the formation of social silos, decreased productivity, increased stress and psychological insecurity, and exposure to a discriminatory environment with limited control over the environment. This data is also found in the literature \cite{de2023benefits}, \cite{de2023post} where it is pointed out that psychological discomfort and exposure to situations that cause psychological suffering and fear for safety permeate the experience of these professionals in face-to-face work.

\section{Implications for Practice}

This study provides actionable insights for fostering inclusivity and improving the developer experience (DX) of LGBTQIAPN+ professionals in agile teams. The findings emphasize the critical role of psychological safety, trust, and open communication in creating inclusive work environments. Organizations can achieve this by investing in diversity and inclusion training, establishing anonymous reporting mechanisms, and strictly enforcing anti-harassment policies. These initiatives not only reduce psychological distress but also cultivate an atmosphere of openness essential for effective collaboration and productivity.

Agile practices, such as retrospectives and pair programming, emerged as valuable tools for building trust and collaboration. Their effectiveness, however, depends on the inclusivity and professionalism of team members and leaders. Companies should adapt these practices to address the unique needs of diverse teams by incorporating mentorship programs and collaborative workshops. Flexible work models, particularly remote and hybrid arrangements, offer significant psychological comfort and productivity benefits, helping LGBTQIAPN+ professionals feel supported while maintaining team cohesion.

The findings also highlight the need to address structural and cultural barriers, such as heteronormative biases and insufficient inclusion policies, which hinder professional growth and engagement. Targeted interventions like diversity quotas, inclusive hiring practices, and leadership development for underrepresented groups are essential to fostering equity and dismantling systemic barriers. Aligning career development opportunities with team and organizational goals through agile ceremonies further supports both individual growth and team cohesion, ensuring diverse teams can thrive in inclusive and high-performing environments.

\section{Conclusion and Future Work}
This study explored the developer experience (DX) of LGBTQIAPN+ professionals in agile teams. Our findings reveal that while agile methodologies emphasize individuals and interactions, structural and cultural barriers—such as discriminatory behaviors and the dominance of heteronormative culture—persist in less mature teams. These challenges undermine trust, collaboration, and psychological safety, hindering team cohesion and productivity. 

This research  highlights the importance of adapting agile practices to the needs of diverse teams, emphasizes the value of flexible work models in promoting inclusivity, and offers actionable strategies for addressing barriers faced by LGBTQIAPN+ professionals. These insights are not only relevant for improving individual and team performance but also for advancing organizational diversity and inclusion efforts.

However, the study's scope is limited to the perceptions of LGBTQIAPN+ professionals in a specific context, and further research is needed to generalize these findings across broader populations. Future studies could investigate the intersectionality of underrepresented groups, assess the scalability of inclusive practices in larger organizations, and explore the long-term impacts of diversity initiatives on agile methodologies.

\section{Threats to validity}

In this section, we examine the types of validity commonly associated with survey research \cite{linaaker2015guidelines} and the reliability of our study. We also outline the mitigation strategies implemented to address threats.

\textbf{Content Validity.} We developed instruments for the opinion survey and refined them through three iterative pilot tests. This iterative approach allowed us to identify and address ambiguities, ensuring that the survey accurately captured the constructs under investigation. Input from subject matter experts was also sought to evaluate and refine the survey questions, aligning them closely with the study’s objectives.

\textbf{External Validity.} It concerns the representativeness of the sample. To address this, we ensured that our sample size of 40 participants exceeded the saturation threshold recommended by Guest et al. \cite{guest2006many}. Furthermore, the Respondent-driven Sampling (RDS) approach was carefully managed to account for potential underrepresentation of isolated individuals by selecting diverse initial seeds and conducting targeted outreach to underrepresented subgroups. We also sent periodic reminders to mitigate nonresponse bias. Clear eligibility criteria and upfront communication were established, ensuring a balanced sample and reducing exclusions.

\textbf{Face Validity.} A potential issue with face validity arises when the survey instrument does not align well with the intended audience. To address this, we carried out a pilot study to assess the instrument's effectiveness. Based on the feedback, we made minor adjustments to enhance its clarity.

\textbf{Internal Validity.} Interpretive validity was a key concern for our study, as it involves the risk of misinterpreting participants’ perspectives. To mitigate this, we paraphrased key statements from open-ended questions to ensure accurate representation of participant responses. The coding process was primarily conducted by the first author, with iterative reviews and contributions from other researchers to refine emerging codes and themes. Additionally, we maintained a detailed audit trail documenting all coding steps, which was shared among the research team and partially made available as supplemental data to ensure transparency.

\textbf{Reliability.} Ensuring the reliability of our findings was critical. To this end, we implemented a structured iterative coding process for analyzing open-ended questions, ensuring consistency and rigor in data interpretation. Moreover, our reliance on pilot tests helped in identifying potential inconsistencies and improving the reliability of the survey instrument. The comprehensive audit trail served as a record of all methodological steps, allowing for reproducibility and validation of our findings.

\bibliographystyle{ieeetr}

\begin{thebibliography}{10}

\bibitem{Grundy2024}
J.~Grundy, T.~Kanij, R.~Hoda, H.~Khalajzadeh, A.~Madugalla, and J.~McIntosh, {\em ED{\&}I and SE: Challenges, Progress, and Lessons}, pp.~17--35.
\newblock Berkeley, CA: Apress, 2024.

\bibitem{de2023benefits}
R.~de~Souza~Santos, C.~V. de~Magalhaes, and P.~Ralph, ``Benefits and limitations of remote work to lgbtqia+ software professionals,'' in {\em 2023 IEEE/ACM 45th International Conference on Software Engineering: Software Engineering in Society (ICSE-SEIS)}, pp.~48--57, IEEE, 2023.

\bibitem{hoffmann2022human}
M.~Hoffmann, D.~Mendez, F.~Fagerholm, and A.~Luckhardt, ``The human side of software engineering teams: an investigation of contemporary challenges,'' {\em IEEE Transactions on Software Engineering}, 2022.

\bibitem{prikladnicki2005aspectos}
R.~Prikladnicki and J.~L.~N. Audy, ``Os aspectos n{\~a}o-t{\'e}cnicos intervenientes no desenvolvimento distribu{\'\i}do de software,'' in {\em Workshop Um Olhar Sociot{\'e}cnico sobre a Engenharia de Software (WOSES)}, vol.~1, pp.~45--55, 2005.

\bibitem{miranda2020towards}
M.~Miranda and R.~Prikladnicki, ``Towards a model for managing diversity and inclusion in software development teams,'' in {\em Proceedings of the 34th Brazilian Symposium on Software Engineering}, pp.~325--331, 2020.

\bibitem{silveira2019systematic}
K.~K. Silveira and R.~Prikladnicki, ``A systematic mapping study of diversity in software engineering: a perspective from the agile methodologies. in 2019 ieee/acm 12th international workshop on cooperative and human aspects of software engineering (chase),'' {\em IEEE, 7{\'s}10}, 2019.

\bibitem{fowler2001agile}
M.~Fowler, J.~Highsmith, {\em et~al.}, ``The agile manifesto,'' {\em Software development}, vol.~9, no.~8, pp.~28--35, 2001.

\bibitem{girardi2021emotions}
D.~Girardi, F.~Lanubile, N.~Novielli, and A.~Serebrenik, ``Emotions and perceived productivity of software developers at the workplace,'' {\em IEEE Transactions on Software Engineering}, vol.~48, no.~9, pp.~3326--3341, 2021.

\bibitem{juarez2021covid}
R.~Ju{\'a}rez-Ram{\'\i}rez, C.~X. Navarro, V.~Tapia-Ibarra, S.~Jim{\'e}nez, C.~Guerra-Garc{\'\i}a, and H.~G. Perez-Gonzalez, ``How covid-19 pandemic affects software developers’ wellbeing: an exploratory study in the west border area of mexico-usa,'' in {\em 2021 9th International conference in software engineering research and innovation (CONISOFT)}, pp.~112--121, IEEE, 2021.

\bibitem{borg2024requirements}
M.~Borg and D.~Graziotin, ``Requirements for organizational resilience: Engineering developer happiness,'' {\em IEEE Software}, vol.~41, no.~4, pp.~14--18, 2024.

\bibitem{graziotin2014happy}
D.~Graziotin, X.~Wang, and P.~Abrahamsson, ``Happy software developers solve problems better: psychological measurements in empirical software engineering,'' {\em PeerJ}, vol.~2, p.~e289, 2014.

\bibitem{graziotin2017consequences}
D.~Graziotin, F.~Fagerholm, X.~Wang, and P.~Abrahamsson, ``Consequences of unhappiness while developing software,'' in {\em 2017 IEEE/ACM 2nd International Workshop on Emotion Awareness in Software Engineering (SEmotion)}, pp.~42--47, IEEE, 2017.

\bibitem{fagerholm2012developer}
F.~Fagerholm and J.~M{\"u}nch, ``Developer experience: Concept and definition,'' in {\em 2012 international conference on software and system process (ICSSP)}, pp.~73--77, IEEE, 2012.

\bibitem{wazlawick2019engenharia}
R.~Wazlawick, {\em Engenharia de software: conceitos e pr{\'a}ticas}.
\newblock Elsevier Editora Ltda., 2019.

\bibitem{de2023post}
R.~de~Souza~Santos, G.~Adisaputri, and P.~Ralph, ``Post-pandemic resilience of hybrid software teams,'' in {\em 2023 IEEE/ACM 16th International Conference on Cooperative and Human Aspects of Software Engineering (CHASE)}, pp.~1--12, IEEE, 2023.

\bibitem{poncell2022diversity}
I.~Poncell and K.~Gama, ``Diversity and inclusion initiatives in brazilian software development companies: Comparing the perspectives of managers and developers,'' in {\em Proceedings of the XXXVI Brazilian Symposium on Software Engineering}, pp.~41--46, 2022.

\bibitem{boman2024breaking}
L.~Boman, J.~Andersson, and F.~G. de~Oliveira~Neto, ``Breaking barriers: Investigating the sense of belonging among women and non-binary students in software engineering,'' in {\em Proceedings of the 46th International Conference on Software Engineering: Software Engineering Education and Training}, pp.~93--103, 2024.

\bibitem{ford2019remote}
D.~Ford, R.~Milewicz, and A.~Serebrenik, ``How remote work can foster a more inclusive environment for transgender developers,'' in {\em 2019 IEEE/ACM 2nd International Workshop on Gender Equality in Software Engineering (GE)}, pp.~9--12, IEEE, 2019.

\bibitem{klotins2023continuous}
E.~Klotins, T.~Gorschek, and M.~Wilson, ``Continuous software engineering: Introducing an industry readiness model,'' {\em IEEE Software}, 2023.

\bibitem{greiler2022actionable}
M.~Greiler, M.-A. Storey, and A.~Noda, ``An actionable framework for understanding and improving developer experience,'' {\em IEEE Transactions on Software Engineering}, vol.~49, no.~4, pp.~1411--1425, 2022.

\bibitem{kropp2020satisfaction}
M.~Kropp, A.~Meier, C.~Anslow, and R.~Biddle, ``Satisfaction and its correlates in agile software development,'' {\em Journal of Systems and Software}, vol.~164, p.~110544, 2020.

\bibitem{dutra2021human}
E.~Dutra, B.~Diirr, and G.~Santos, ``Human factors and their influence on software development teams-a tertiary study,'' in {\em Proceedings of the XXXV Brazilian Symposium on Software Engineering}, pp.~442--451, 2021.

\bibitem{machuca2022perceptions}
L.~Machuca-Villegas, G.~P. Gasca-Hurtado, S.~M. Puente, and L.~M.~R. Tamayo, ``Perceptions of the human and social factors that influence the productivity of software development teams in colombia: A statistical analysis,'' {\em Journal of systems and software}, vol.~192, p.~111408, 2022.

\bibitem{d2024measuring}
S.~D’Angelo, J.~Lin, J.~Dicker, C.~Egelman, M.~Hodges, C.~Green, and C.~Jaspan, ``Measuring developer experience with a longitudinal survey,'' {\em IEEE Software}, vol.~41, no.~4, pp.~19--24, 2024.

\bibitem{razzaq2024systematic}
A.~Razzaq, J.~Buckley, Q.~Lai, T.~Yu, and G.~Botterweck, ``A systematic literature review on the influence of enhanced developer experience on developers' productivity: Factors, practices, and recommendations,'' {\em ACM Computing Surveys}, vol.~57, no.~1, pp.~1--46, 2024.

\bibitem{rodriguez2021perceived}
G.~Rodr{\'\i}guez-P{\'e}rez, R.~Nadri, and M.~Nagappan, ``Perceived diversity in software engineering: a systematic literature review,'' {\em Empirical Software Engineering}, vol.~26, pp.~1--38, 2021.

\bibitem{de2020diversity}
N.~P.~R. de~Souza and K.~Gama, ``Diversity and inclusion: Culture and perception in information technology companies,'' {\em IEEE Revista Iberoamericana de Tecnologias del Aprendizaje}, vol.~15, no.~4, pp.~352--361, 2020.

\bibitem{prana2021including}
G.~A.~A. Prana, D.~Ford, A.~Rastogi, D.~Lo, R.~Purandare, and N.~Nagappan, ``Including everyone, everywhere: Understanding opportunities and challenges of geographic gender-inclusion in oss,'' {\em IEEE Transactions on Software Engineering}, vol.~48, no.~9, pp.~3394--3409, 2021.

\bibitem{duran2024exploring}
A.~Dur{\'a}n~Toro, P.~Fern{\'a}ndez, B.~Bern{\'a}rdez, N.~Weinman, A.~Akal{\i}n, and A.~Fox, ``Exploring gender bias in remote pair programming among software engineering students: The twincode original study and first external replication,'' {\em Empirical Software Engineering}, vol.~29, no.~2, 2024.

\bibitem{imtiaz2019investigating}
N.~Imtiaz, J.~Middleton, J.~Chakraborty, N.~Robson, G.~Bai, and E.~Murphy-Hill, ``Investigating the effects of gender bias on github,'' in {\em 2019 IEEE/ACM 41st International Conference on Software Engineering (ICSE)}, pp.~700--711, IEEE, 2019.

\bibitem{kanij2024enhancing}
T.~Kanij, J.~Grundy, and J.~McIntosh, ``Enhancing understanding and addressing gender bias in it/se job advertisements,'' {\em Journal of Systems and Software}, vol.~217, p.~112169, 2024.

\bibitem{trinkenreich2022women}
B.~Trinkenreich, I.~Wiese, A.~Sarma, M.~Gerosa, and I.~Steinmacher, ``Women’s participation in open source software: A survey of the literature,'' {\em ACM Transactions on Software Engineering and Methodology (TOSEM)}, vol.~31, no.~4, pp.~1--37, 2022.

\bibitem{weisshaar2024hiring}
K.~Weisshaar, K.~Chavez, and T.~Hutt, ``Hiring discrimination under pressures to diversify: Gender, race, and diversity commodification across job transitions in software engineering,'' {\em American Sociological Review}, vol.~89, no.~3, pp.~584--613, 2024.

\bibitem{hussain2020human}
W.~Hussain, H.~Perera, J.~Whittle, A.~Nurwidyantoro, R.~Hoda, R.~A. Shams, and G.~Oliver, ``Human values in software engineering: Contrasting case studies of practice,'' {\em IEEE Transactions on Software Engineering}, vol.~48, no.~5, pp.~1818--1833, 2020.

\bibitem{trinkenreich2022empirical}
B.~Trinkenreich, R.~Britto, M.~A. Gerosa, and I.~Steinmacher, ``An empirical investigation on the challenges faced by women in the software industry: A case study,'' in {\em Proceedings of the 2022 ACM/IEEE 44th International Conference on Software Engineering: Software Engineering in Society}, pp.~24--35, 2022.

\bibitem{campero2021hiring}
S.~Campero, ``Hiring and intra-occupational gender segregation in software engineering,'' {\em American Sociological Review}, vol.~86, no.~1, pp.~60--92, 2021.

\bibitem{canedo2021breaking}
E.~D. Canedo, F.~Mendes, A.~Cerqueira, M.~Okimoto, G.~Pinto, and R.~Bonifacio, ``Breaking one barrier at a time: how women developers cope in a men-dominated industry,'' in {\em Proceedings of the XXXV Brazilian Symposium on Software Engineering}, pp.~378--387, 2021.

\bibitem{van2023still}
S.~Van~Breukelen, A.~Barcombt, S.~Baltes, and A.~Serebrenik, ``“still around”: Experiences and survival strategies of veteran women software developers,'' in {\em 2023 IEEE/ACM 45th International Conference on Software Engineering (ICSE)}, pp.~1148--1160, IEEE, 2023.

\bibitem{gama2024much}
K.~Gama, A.~P. Chaves, D.~M. Ribeiro, K.~Devathasan, and D.~Damian, ``How much do you know about your users? a study of developer awareness about diverse users,'' in {\em 2024 IEEE 32nd International Requirements Engineering Conference Workshops (REW)}, pp.~110--118, IEEE, 2024.

\bibitem{dagan2023building}
E.~Dagan, A.~Sarma, A.~Chang, S.~D’Angelo, J.~Dicker, and E.~Murphy-Hill, ``Building and sustaining ethnically, racially, and gender diverse software engineering teams: A study at google,'' in {\em Proceedings of the 31st ACM Joint European Software Engineering Conference and Symposium on the Foundations of Software Engineering}, pp.~631--643, 2023.

\bibitem{albusays2021diversity}
K.~Albusays, P.~Bjorn, L.~Dabbish, D.~Ford, E.~Murphy-Hill, A.~Serebrenik, and M.-A. Storey, ``The diversity crisis in software development,'' {\em IEEE Software}, vol.~38, no.~2, pp.~19--25, 2021.

\bibitem{sanchez2021framework}
M.~S{\'a}nchez-Gord{\'o}n and R.~Colomo-Palacios, ``A framework for intersectional perspectives in software engineering,'' in {\em 2021 IEEE/ACM 13th International Workshop on Cooperative and Human Aspects of Software Engineering (CHASE)}, pp.~121--122, IEEE, 2021.

\bibitem{richard2021effects}
O.~C. Richard, M.~d.~C. Triana, and M.~Li, ``The effects of racial diversity congruence between upper management and lower management on firm productivity,'' {\em Academy of Management Journal}, vol.~64, no.~5, pp.~1355--1382, 2021.

\bibitem{gunay2020improving}
C.~G{\"u}nay, A.~Doloc-Mihu, R.~Barakat, T.~Gluick, and C.~A. Moore, ``Improving critical thinking in software development via interdisciplinary projects at a most diverse college,'' in {\em Proceedings of the 21st Annual Conference on Information Technology Education}, pp.~206--212, 2020.

\bibitem{verwijs2023double}
C.~Verwijs and D.~Russo, ``The double-edged sword of diversity: How diversity, conflict, and psychological safety impact software teams,'' {\em IEEE Transactions on Software Engineering}, 2023.

\bibitem{santos2024exploring}
R.~d.~S. Santos, C.~Magalhaes, R.~Santos, and J.~Correia-Neto, ``Exploring hybrid work realities: A case study with software professionals from underrepresented groups,'' in {\em Companion Proceedings of the 32nd ACM International Conference on the Foundations of Software Engineering}, pp.~27--37, 2024.

\bibitem{ezeilo2023coronovirus}
C.~O. Ezeilo and J.~Green-McKenzie, ``The coronovirus-19 pandemic and the future of work,'' {\em Journal of Occupational and Environmental Medicine}, pp.~10--1097, 2023.

\bibitem{hsu2024remote}
D.~H. Hsu and P.~B. Tambe, ``Remote work and job applicant diversity: Evidence from technology startups,'' {\em Management Science}, 2024.

\bibitem{nicholson2022remote}
S.~Nicholson, G.~Popoola, M.~McKie, J.~Moten, and T.~Fletcher, ``Remote work and satisfaction for black engineers and computer scientists,'' in {\em 2022 IEEE Frontiers in Education Conference (FIE)}, pp.~1--7, IEEE, 2022.

\bibitem{li2024you}
Z.~S. Li, D.~Ly, L.~Nagel, N.~N. Arony, and D.~Damian, ``“do you have time for a quick call?”: Exploring remote and hybrid requirements engineering practices and challenges in industry,'' in {\em 2024 IEEE 32nd International Requirements Engineering Conference (RE)}, pp.~43--54, IEEE, 2024.

\bibitem{johnson2024strategies}
B.~Johnson, ``Strategies for reporting and centering marginalized developer experiences,'' {\em Equity, Diversity, and Inclusion in Software Engineering}, p.~507, 2024.

\bibitem{molleri2016survey}
J.~S. Moll{\'e}ri, K.~Petersen, and E.~Mendes, ``Survey guidelines in software engineering: An annotated review,'' in {\em Proceedings of the 10th ACM/IEEE international symposium on empirical software engineering and measurement}, pp.~1--6, 2016.

\bibitem{basili1994gqm}
V.~Basili, ``Gqm approach has evolved to include models,'' {\em IEEE SOFTWARE}, vol.~11, no.~1, pp.~8--8, 1994.

\bibitem{katzenbach2001equipes}
J.~R. Katzenbach and D.~K. SMiTH, {\em Equipes de alta performance: conceitos, princ{\'\i}pios e t{\'e}cnicas para potencializar o desempenho das equipes}.
\newblock Gulf Professional Publishing, 2001.

\bibitem{liboreiro2018gestao}
K.~R. Liboreiro, R.~S. Guimar{\~a}es, {\em et~al.}, ``Gest{\~a}o de equipes de alto desempenho: abordagens e discuss{\~o}es recentes,'' {\em Gest{\~a}o \& Regionalidade}, vol.~34, no.~102, pp.~5--22, 2018.

\bibitem{montanari2011maturidade}
R.~L. Montanari, L.~A. Pilatti, I.~A.~d. Lima, and C.~A. Romano, ``A maturidade e o desempenho das equipes no ambiente produtivo,'' {\em Gest{\~a}o \& Produ{\c{c}}{\~a}o}, vol.~18, pp.~367--378, 2011.

\bibitem{santos2014social}
D.~d. A. F.~d. Santos, L.~Mour{\~a}o, and L.~A.~M. Naiff, ``Social representations about teamwork,'' {\em Psicologia: Ci{\^e}ncia e Profiss{\~a}o}, vol.~34, no.~3, p.~643, 2014.

\bibitem{baltesandralph}
S.~Baltes and P.~Ralph, ``Sampling in software engineering research: A critical review and guidelines,'' {\em Empirical Software Engineering}, vol.~27, no.~4, p.~94, 2022.

\bibitem{de2024hidden}
R.~de~Souza~Santos and K.~Gama, ``Hidden populations in software engineering: Challenges, lessons learned, and opportunities,'' in {\em Proceedings of the 1st IEEE/ACM International Workshop on Methodological Issues with Empirical Studies in Software Engineering}, pp.~58--63, 2024.

\bibitem{cruzes2011recommended}
D.~S. Cruzes and T.~Dyba, ``Recommended steps for thematic synthesis in software engineering,'' in {\em 2011 international symposium on empirical software engineering and measurement}, pp.~275--284, IEEE, 2011.

\bibitem{roberts2024artificial}
J.~Roberts, M.~Baker, and J.~Andrew, ``Artificial intelligence and qualitative research: The promise and perils of large language model (llm)‘assistance’,'' {\em Critical Perspectives on Accounting}, vol.~99, p.~102722, 2024.

\bibitem{yan2024human}
L.~Yan, V.~Echeverria, G.~M. Fernandez-Nieto, Y.~Jin, Z.~Swiecki, L.~Zhao, D.~Ga{\v{s}}evi{\'c}, and R.~Martinez-Maldonado, ``Human-ai collaboration in thematic analysis using chatgpt: A user study and design recommendations,'' in {\em Extended Abstracts of the CHI Conference on Human Factors in Computing Systems}, pp.~1--7, 2024.

\bibitem{SMITE2021106612}
D.~Šmite, N.~B. Moe, and J.~Gonzalez-Huerta, ``Overcoming cultural barriers to being agile in distributed teams,'' {\em Information and Software Technology}, vol.~138, p.~106612, 2021.

\bibitem{sarker2022identification}
J.~Sarker, ``Identification and mitigation of toxic communications among open source software developers,'' in {\em Proceedings of the 37th IEEE/ACM International Conference on Automated Software Engineering}, pp.~1--5, 2022.

\bibitem{riaz2019effect}
M.~N. Riaz, A.~Buriro, and A.~Mahboob, ``The effect of software development project team structure on the process of knowledge sharing: An empirical study,'' in {\em 2019 2nd International Conference on Computing, Mathematics and Engineering Technologies (iCoMET)}, pp.~1--5, IEEE, 2019.

\bibitem{gunawardena2022destructive}
S.~D. Gunawardena, P.~Devine, I.~Beaumont, L.~P. Garden, E.~Murphy-Hill, and K.~Blincoe, ``Destructive criticism in software code review impacts inclusion,'' {\em Proceedings of the ACM on Human-Computer Interaction}, vol.~6, no.~CSCW2, pp.~1--29, 2022.

\bibitem{trinkenreich2021please}
B.~Trinkenreich, ``Please don't go—a comprehensive approach to increase women's participation in open source software,'' in {\em 2021 IEEE/ACM 43rd International Conference on Software Engineering: Companion Proceedings (ICSE-Companion)}, pp.~293--298, IEEE, 2021.

\bibitem{meyer2014software}
A.~N. Meyer, T.~Fritz, G.~C. Murphy, and T.~Zimmermann, ``Software developers' perceptions of productivity,'' in {\em Proceedings of the 22nd ACM SIGSOFT International Symposium on Foundations of Software Engineering}, pp.~19--29, 2014.

\bibitem{ahmad2024non}
M.~O. Ahmad, T.~Gustavsson, A.~Katin, N.~Tau{\v{s}}an, and V.~Mandi{\'c}, ``Non-technical aspects of technical debt in the context of large-scale agile development: A qualitative study,'' in {\em 2024 50th Euromicro Conference on Software Engineering and Advanced Applications (SEAA)}, pp.~260--267, IEEE, 2024.

\bibitem{graziotin2017unhappy}
D.~Graziotin, F.~Fagerholm, X.~Wang, and P.~Abrahamsson, ``Unhappy developers: Bad for themselves, bad for process, and bad for software product,'' in {\em 2017 IEEE/ACM 39th International Conference on Software Engineering Companion (ICSE-C)}, pp.~362--364, IEEE, 2017.

\bibitem{steinmacher2024breaking}
I.~Steinmacher, B.~Trinkenreich, A.~Sarma, and M.~Gerosa, ``Breaking the glass floor for women in tech,'' in {\em Equity, Diversity, and Inclusion in Software Engineering: Best Practices and Insights}, pp.~55--66, Apress Berkeley, CA, 2024.

\bibitem{do2023investigating}
J.~C.~S. do~Out{\~a}o, L.~A.~M. da~Costa, R.~P. dos Santos, and A.~Serebrenik, ``Investigating the barriers that women face in software development teams focusing on the context of proprietary software ecosystems,'' in {\em International Conference on Software Business}, pp.~164--170, Springer Nature Switzerland Cham, 2023.

\bibitem{meyer2019today}
A.~N. Meyer, E.~T. Barr, C.~Bird, and T.~Zimmermann, ``Today was a good day: The daily life of software developers,'' {\em IEEE Transactions on Software Engineering}, vol.~47, no.~5, pp.~863--880, 2019.

\bibitem{fontao2023developer}
A.~Font{\~a}o, S.~Cleger-Tamayo, I.~Wiese, R.~Pereira~dos Santos, and A.~Claudio Dias-Neto, ``A developer relations (devrel) model to govern developers in software ecosystems,'' {\em Journal of Software: Evolution and Process}, vol.~35, no.~5, p.~e2389, 2023.

\bibitem{linaaker2015guidelines}
J.~Lin{\aa}ker, S.~M. Sulaman, R.~M. de~Mello, and M.~H{\"o}st, ``Guidelines for conducting surveys in software engineering,'' 2015.

\bibitem{guest2006many}
G.~Guest, A.~Bunce, and L.~Johnson, ``How many interviews are enough? an experiment with data saturation and variability,'' {\em Field methods}, vol.~18, no.~1, pp.~59--82, 2006.

\end{thebibliography}

\end{document}